# On the Origin of Preferential Growth of Semiconducting Single-Walled Carbon Nanotubes

Yiming Li, Shu Peng[†], David Mann, Jien Cao, Ryan Tu, K. J. Cho[†] and Hongjie Dai*

Department of Chemistry and Laboratory for Advanced Materials, Stanford University, Stanford, CA 94305, USA

[†] Department of Mechanical Engineering, Stanford University, Stanford, CA 94305, USA

A correlation is observed between the diameter ($d$) distribution of single walled carbon nanotubes and the percentages of metallic and semiconducting tubes in materials synthesized at low temperature (600 ºC) by plasma-assisted chemical vapor deposition. Small diameter nanotubes (average $d$~1.1 nm) show semiconducting-tube percentage much higher than expected for random chirality distribution. Density functional theory calculations reveal discernable differences in the cohesive energies and heat of formation energies for similar-diameter metallic, quasi-metallic and semiconducting nanotubes. Semiconducting nanotubes exhibit the lowest energies and the stabilization effect scales with ~$1/d^2$. This is a likely thermodynamic factor in preferential growth of small diameter semiconducting nanotubes.

* Email: hdai@stanford.edu

The chirality of a single-walled carbon nanotube (SWNT) determines whether the nanotube is a metal or semiconductor[1]. While semiconducting SWNTs (S-SWNTs) can be used to build high performance field effect transistors and sensors[2-4], metallic SWNTs (M-SWNTs) might be useful for interconnects. It is clear however, that these potential applications will hinge upon obtaining purely metallic and semiconducting nanotubes[5]. One approach might be chemically separating metallic from semiconducting SWNTs[6-10], and the other could be via selective growth to preferentially produce a certain type of nanotubes[11]. Recently, we reported a plasma-enhanced chemical vapor deposition (PECVD) method producing SWNTs at 600°C with the percentage of S-SWNTs ~89%[11]. In other work, with materials synthesized by a thermal CVD method, Bachilo et al. even observed that SWNTs with certain chiralities were preferentially formed[12]. It appears that selective growth of certain types of nanotubes can indeed occur. However, little is known thus far about the driving forces behind the preferential growth phenomena.

Here, we report that higher percentages of S-SWNTs are grown when the growth parameters are adjusted to produce smaller diameter SWNTs at a relatively low growth temperature of 600 ºC. First principles calculations reveal that SWNTs of different metallicity exhibit discernable differences in the heat of formation energies. S-SWNTs show the lowest energies and are the most stable compared to metallic and quasi-metallic[13] SWNTs (QM-SWNT with band gaps ~ tens of meV). This could be a factor leading to the preferential growth of semiconducting nanotubes.

We carried out SWNTs synthesis in a 4-inch quasi-remote PECVD system (RF, 13.56 MHz) at 600 °C using discrete ferritin nanoparticles[14] as catalysts on $SiO_2$ (thickness $t_{ox}$=67nm)/Si substrates as described previously[11]. The methane flow rate was 60 sccm (standard cubic centimeter per minute) and the pressure in the PECVD system was maintained at 500 mTorr. The plasma power was varied from 50W to 200W and the growth time was typically 3 minutes.

We found that at a fixed growth temperature of 600 °C, varying the plasma power from 50 W to 200 W led to an obvious shift in the diameter distribution of the synthesized nanotubes. Atomic force microscopy (AFM) imaging (Fig. 1a, b) and topographic measurements (Fig. 1c, d) revealed the mean diameters of nanotubes produced by 50W and 200W PECVD are $<d>=1.12 \pm 0.25$nm and $<d>=1.66 \pm 0.38$nm respectively. More than 80% of the nanotubes in the 50W sample (referred to as *sample-A* herein) had $d<1.4$nm (Fig. 1e) while most (~76%) nanotubes in the 200W sample (*sample-B*) exhibited $d>1.4$nm (Fig. 1f). Note that the catalysts used for *sample-A* and *B* were the same with diameters in the range of 0.7 to 3 nm. Under the otherwise identical growth conditions, higher plasma power apparently allowed for the growth of larger diameter SWNTs from larger particles as evidenced by the clear shift in the SWNT diameter distribution. We attribute this to more effective decomposition of methane at higher plasma power providing more efficient carbon feedstock needed for super-saturation of larger catalyst particles and thus growth of larger tubes.

With the synthesized nanotubes on $SiO_2$/Si substrates, we fabricated large arrays of three-terminal devices each comprised of a Pd[3] source (S), drain (D), a back-Si-gate and one or multiple SWNTs bridging S/D (Fig. 2a). We then used electrical transport measurements to assess whether the SWNTs are metallic or semiconducting[11]. Devices comprised of only S-SWNTs exhibited several orders of magnitude conductance changes ('depletable') under electrostatic gating (Fig. 2b), whereas devices comprised of M-and/or QM-SWNTs exhibit weak gate dependence ('non-depletable' for both M- and QM-SWNTs). We used an electrical breakdown (Fig. 2c) method detailed previously [11] to 'count' larger numbers of M-/QM- and S-SWNTs in each device. Using this method, we identified ~85% and ~75% S-SWNTs in *sample-A* and –B respectively (Table 1). The former was similar to that of ~ 89% S-SWNTs in samples synthesized under similar conditions (i.e., low plasma power). These results reveal preferential growth of S-SWNTs in small-diameter ($<d>$=~1.1 nm) samples and reduced percentages of S-SWNTs in the larger-diameter sample ($<d>$= ~ 1.7 nm).

Fig. 3a and 3b summarize the maximum device conductance ($G_{max}$ measured over the experimentally accessible gate-voltage range) measured with over 100 devices for *sample-A* and *-B* respectively. We observe that the depletable devices (comprised of only S-SWNTs) generally exhibit lower $G_{max}$ than the non-depletable devices bridged by M- or QM-SWNTs. This is attributed to differences in junction resistance due to the existence of Schottky barriers (SBs) at the metal/S-SWNT contacts and no significant SBs at the contacts for M/QM-SWNTs. Large numbers of the depletable devices in *sample-B* (Fig. 3b) exhibit higher $G_{max}$ than those in *sample-A* (Fig. 3a), corresponding to the larger diameter S-SWNTs in *sample-B* affording lower SBs at the Pd contacts due to smaller band gaps ($E_{gap} \sim 1/d$) and thus higher ON-conductance[3]. Importantly, for both *sample-A* and *–B*, devices with lower $G_{max}$, i.e. comprised of smaller diameter tubes clearly show higher percentages of depletable devices than non-depletable ones. This can be gleaned from Fig.3 that the ratio between light (percentage of depletable devices) and dark (non-depletable devices) bars in each $G_{max}$ range tends to be larger for smaller $G_{max}$ or smaller diameter tubes. This again suggests higher percentages of S-SWNTs for tubes with smaller diameters.

To understand the origin of preferential formation of S-SWNTs, we carried out density functional theory (DFT) calculations to investigate the energetics and stability of various types of nanotubes. Self-consistent electronic structure calculations were performed using the ultra-soft pseudo-potential code VASP[15] with local density approximation (LDA) at a kinetic energy cutoff of 21 Rydberg. Super-cells containing one-unit cells of nanotubes (number of C-atoms in unit cell depends on diameter and chirality and is up to 196) were used. The Brillouin zone sampling was approximated by 12 k-points along the tube axis. Electronic minimization was carried out to a convergence tolerance of 0.1 meV and the structures were relaxed until the maximum forces on any carbon atom was less than 0.015 eV/Å. The cohesive energy and heat of formation of SWNT per C-atom is defined as:

$$\Delta H = E_{total}(SWNT)/(number\ of\ atoms) - E(C) \qquad (1)$$

where $E(C)$ is the energy for an isolated C-atom. As examples, Fig. 4a shows the density of states for (11,11), (18,0) and (19,0) M, QM and S-SWNTs with d= 1.49 nm, 1.40 nm and 1.49 nm and band gap $E_g$ = 0, 90 meV and 0.55 eV respectively. The atomic heat of formation energies for various M-, QM- and SWNTs with diameters in the range of $d$= 1 to 2 nm together with that of a flat graphene are plotted in Fig. 4b. The heat of formation energy for the three types of tubes relative to that of graphene can be well fit into $\sim 1/d^2$ forms (solid lines in Fig. 4b). Importantly, for a given $d$, we consistently find that S-SWNTs exhibit lower formation energies than QM-SWNTs by ~ 4 (meV)/$d^2$ (Fig. 4c solid line) and than M-SWNTs by ~ 9 (meV)/$d^2$ (Fig. 4c dashed line).

The $\Delta H - \Delta H(graphene) \sim 1/d^2$ fits in Fig. 4b are in agreement with theoretical calculations performed previously[16-18] and are mainly due to strain energies in the tubes that are absent in a flat graphene. The discernable differences in the formation energies for metallic, quasi-metallic and semiconducting SWNTs are revealed here for the first time, as earlier calculations have mainly elucidated the diameter dependence but not the chirality or metallicity dependence. It is reasonable that for similar diameter, S-SWNTs are lower in energy than metallic tubes due to electronic energy gain resulted from band gap opening in the former. A relevant topic is that the driving force for Peierls instability in a one-dimensional (1D) metal is band gap opening for lowering the electronic energy of the system. Metallic quasi-1D SWNTs do exist since they are stable against Peierls distortion over a wide range of temperatures due to the high cost in lattice distortion energy[19]. Nevertheless, M-SWNTs do exhibit higher electronic energies than semiconducting tubes with similar diameter.

The lower energy of S-SWNTs than M/QM-SWNTs is more appreciable for nanotubes with smaller diameters than larger ones due to the $\sim 1/d^2$ dependence (Fig.4). This theoretical result may be related to the preferential formation of semiconducting nanotubes (for tubes with d ~1.1 nm) and suggests that the energies and thermodynamics associated with the various nanotube structures could be a factor in determining the types of nanotubes grown. The relatively low temperature used in our SWNT synthesis may allow for a clear manifestation of the preferential growth effect. We did vary the PECVD growth temperature from 600°C to 750°C and observed no drastic differences in the diameter distribution and percentage of S-SWNTs for sample grown in this temperature range. For even higher temperatures of 850°C -900°C of our typical methane CVD[14,20], the same catalyst afforded nanotubes with d=1.85 ± 0.55 nm without any preference in semiconducting tubes (percentage ~ 60 to 70%).

It is instructive to note that for S- and M-SWNTs with $d$~1nm, the per unit length formation energy is lower for S-SWNTs by ~0.15eV per C-C bond length of 1.46 Å (or a stabilization energy of ~ 1.0eV/nm) based on the atomic heat of formation energies in Fig. 4. This is substantial ($k_B$T~0.07eV at 600 ºC) and could be a driving force for lengthening of nanotubes from the seed catalyst particles[14] preferential into semiconductors. We do note that energetics is certainly only one of the factors involved in the growth process. The fact that at a fixed 600 ºC, increasing carbon feedstock by higher plasma power allowed for larger particles producing larger diameter SWNTs clearly suggests the importance of the chemistry in the nanoparticles and the dynamic and kinetic factors involved. Our results here do reveal that in the temperature range investigated, the small diameter SWNTs produced tend to be rich in semiconductors.

Our theoretical results here may be in agreement with the higher chemical functionalization reactivity for M-SWNTs than S-SWNTs[21], a phenomenon reported recently and potentially useful for separation of various types of nanotubes. Another note is that it remains to be seen whether it holds true in materials synthesized by other methods that the percentage of S-SWNTs is correlated with nanotube diameter distribution. We do note a recent report of an unexplained high enrichment of S-SWNTs for nanotube fraction with small diameters (< ~1nm) probed by Raman spectroscopy in a separation work with the Hipco material[7]. Such enrichment could be related to the existence of higher percentage of small diameter S-SWNTs in the starting as-grown material. While many growth issues remain to be understood, our current work makes a step forward towards such understanding. It is clear that better control of catalyst particles with a truly narrow distribution and the development of low temperature growth process will allow for better control of the homogeneity of the synthesized nanotubes.

**Acknowledgement.** This work was supported by Intel and MARCO MSD Focus Center. We acknowledge helpful discussions with Marco Rolandi.

**Figure Captions**

**Figure 1.** (a) AFM image of nanotubes grown by PECVD with a plasma power of 50W (*sample-A*). (b) AFM image of nanotubes grown with a plasma power of 200W (*sample-B*). The z-ranges for both images are 10 nm. Tubes in (b) appear brighter (larger in diameter) than those in (a). (c) A topographic height profile along the line drawn in (a). (d) A topographic height profile along the line drawn in (b). (e) Diameter distribution for *sample-A* obtained over 124 tubes. (f) Diameter distribution for *sample-B* obtained over 167 tubes.

**Figure 2.** (a) An AFM image of two nanotubes bridging Pd S/D electrodes in a device (d ≈ 1.2 nm and 1.3nm for the two tubes respectively, channel length L~300 nm). The two tubes were located several microns apart on the substrate and the region in between is not shown in the image for clarity. (b) Current vs. gate-voltage ($I_{ds}$ vs. $V_g$) curve (recorded under a bias of $V_{ds}$=100 mV) for the 2-tube device showing $10^6$ p-channel conductance depletion by sweeping the gate. Both tubes in the device are semiconductors. (b) $I_{ds}$ vs. $V_{ds}$ recorded under $V_g$=-5V (both tubes in ON state) showing sequential electrical breakdown of the two tubes in the device.

**Table I.** Diameter distributions, number of SWNTs analyzed, and percentages of S- and M/QM-SWNTs for two samples synthesized using two different plasma power. The error bars for the percentages are calculated using $1.96[p(1-p)/N_T]^{1/2}$ with confidence level of 95% for proportionate sampling where $N_T$ is the total number of SWNTs analyzed and p is the mean percentage value.

**Figure 3**. (a) and (b), for *sample-A* and *sample-B* respectively, the graph shows the percentages of depletable and non-depletable devices with $G_{max}$ (i.e., maximum conductance measured in the –5 to 5 V gate voltage range) in various ranges (x-axis).

**Figure 4.** (a) Density of states vs. energy for three M-, QM- and S-SWNTs respectively from DFT calculations. (b) DFT calculations of atomic heat of formation vs. tube diameter $d$ for various M-, QM- and S-SWNTs (symbols). The solid lines are $\sim 1/d^2$ fits for the formation energy relative to graphene for the three types of tubes. (c) Difference in formation energies between M- and S-SWNTs (dashed line) and QM- and S-SWNTs (solid line) respectively vs. tube diameter.

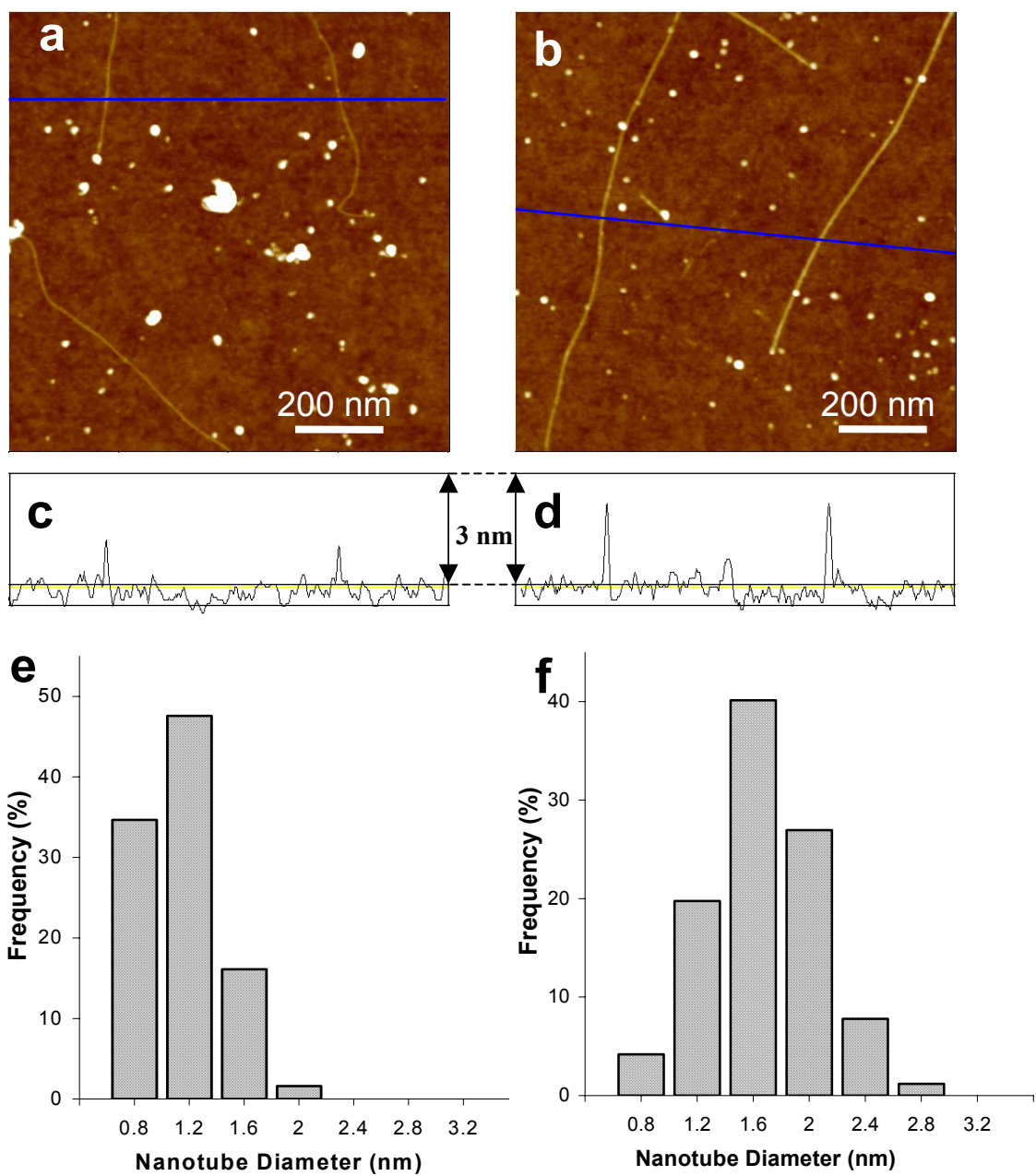

Fig 1

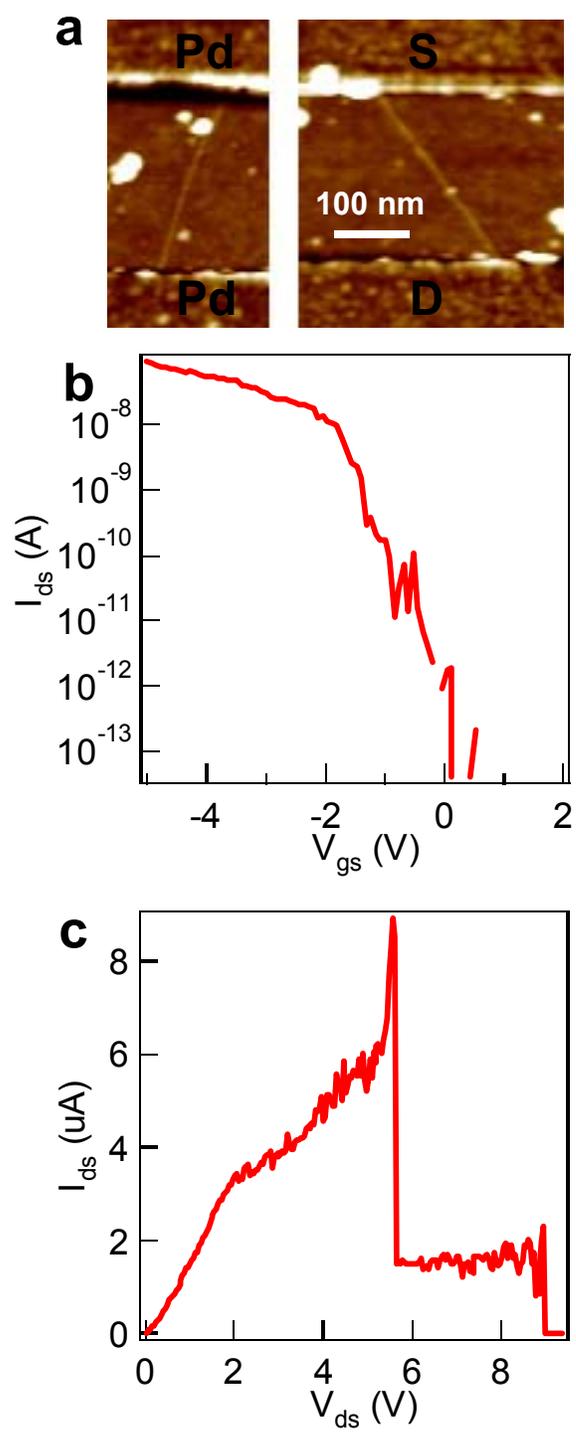

|  | Samples | SWNT diameter | total # of devices | total # of SWNTs | # of S-SWNTs | # of M/QM-SWNTs | S-SWNT percentage |
|---|---|---|---|---|---|---|---|
| **PECVD 600 ºC** | *Sample-A* (50W plasma) | 1.12 ± 0.25 nm | 226 | 386 | 331 | 55 | 85.6% ± 3.5% |
| | *Sample-B* (200W plasma) | 1.66 ± 0.38 nm | 212 | 335 | 252 | 83 | 75.2% ± 4.6% |

Table 1

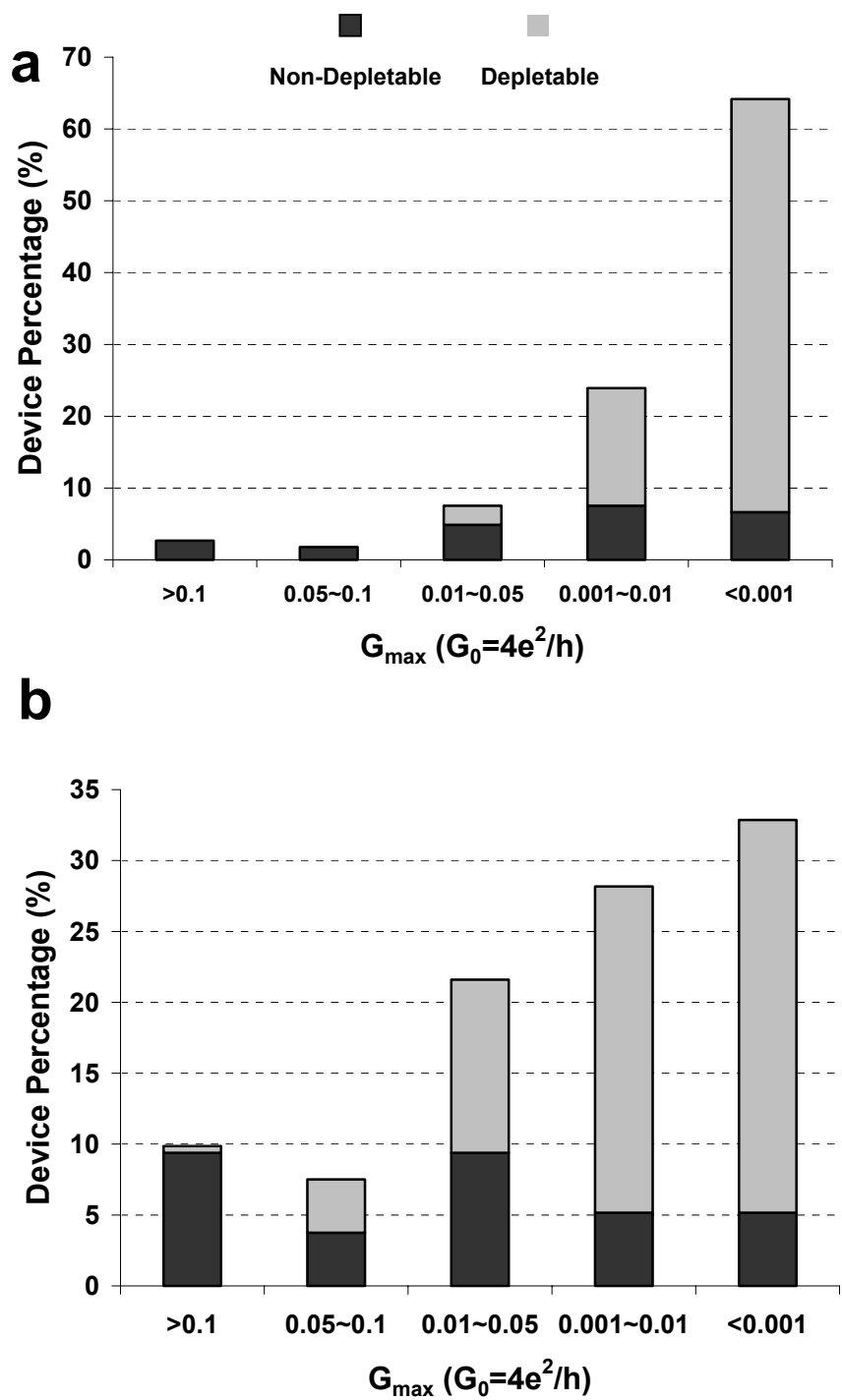

Fig 3

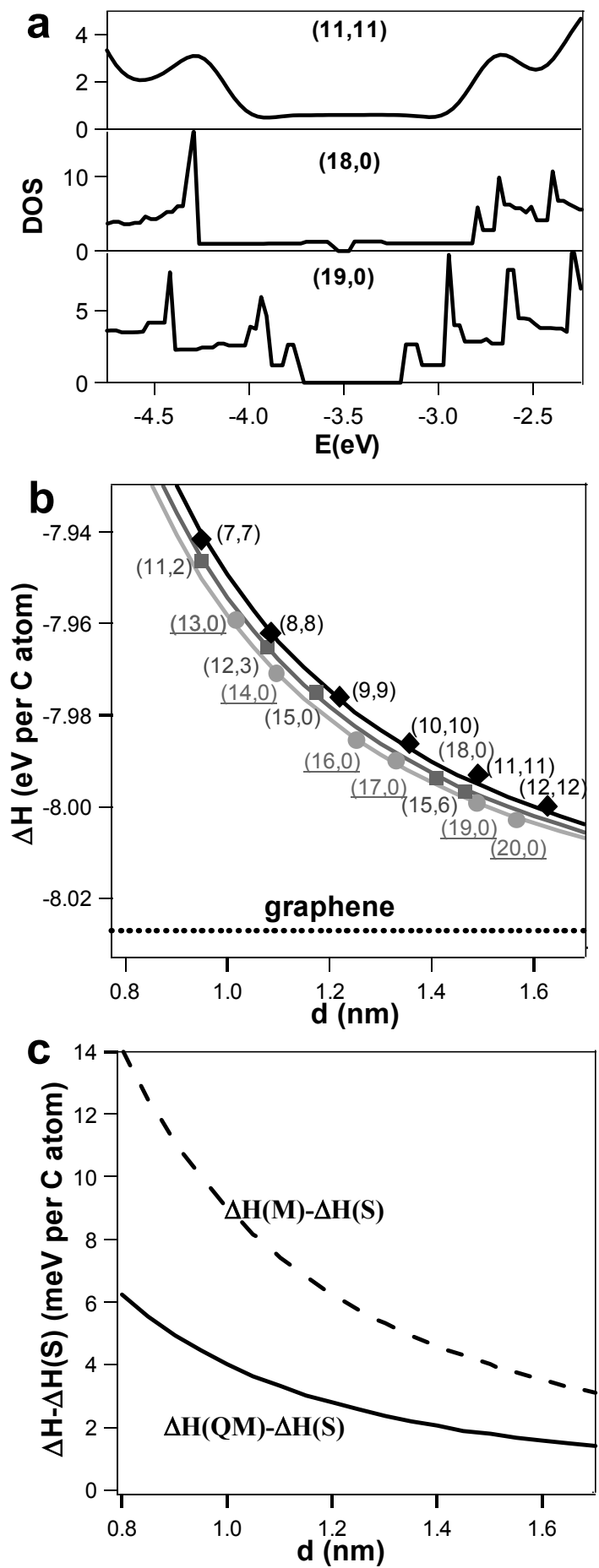

Fig 4